\documentclass[10pt,conference]{IEEEtran}
\usepackage{mathtools}

\usepackage{lipsum}
\usepackage{tikz}     
\usepackage{pifont} 

\usepackage{booktabs}
\usepackage{siunitx}

\usepackage{amsmath}
\usepackage{amsfonts,amssymb,amsbsy}

\IEEEoverridecommandlockouts        
\usepackage{graphics}
\usepackage{epsfig} 
\usepackage{siunitx} 
\usepackage{array}   

\usepackage{xcolor} 
\usepackage{caption}

\usepackage{mathptmx} 
\usepackage{times} 
\usepackage{array}
\usepackage{hyperref}
\usepackage{stmaryrd}  

\usepackage{blindtext, graphicx}   
\usepackage[tc]{titlepic}   
\usepackage{float}   
\usepackage{cuted}     
\usepackage{amsthm}   
\newtheoremstyle{case}{}{}{}{}{}{:}{ }{}
\usepackage{authblk}
\usepackage{caption}
\usepackage{subcaption}

\usepackage{xcolor}
\usepackage{soul}
\newcommand{\comment}[1]{}

\begin{document}

\title{Enhancing Open-World Bacterial Raman Spectra Identification by Feature Regularization for Improved Resilience against Unknown Classes}

\date{\vspace{-5ex}}

\author[*,1]{Yaroslav Balytskyi}
\author[2]{Nataliia Kalashnyk}
\author[3]{Inna Hubenko}
\author[3]{Alina Balytska}
\author[4]{Kelly McNear}

\affil[1]{Department of Physics and Astronomy, Wayne State University, Detroit, MI, 48201, USA}

\affil[2]{National University of Civil Protection of Ukraine, Cherkasy, 18034, Ukraine}

\affil[3]{Cherkasy Medical Academy, Cherkasy, 18000, Ukraine}

\affil[4]{UCCS BioFrontiers Center, University of Colorado, Colorado Springs, Colorado, 80918, USA}

\maketitle

\begin{abstract}

The combination of Deep Learning techniques and Raman spectroscopy shows great potential offering precise and prompt identification of pathogenic bacteria in clinical settings. However, the traditional closed-set classification approaches assume that all test samples belong to one of the known pathogens, and their applicability is limited since the clinical environment is inherently unpredictable and dynamic, unknown or emerging pathogens may not be included in the available catalogs. We demonstrate that the current state-of-the-art Neural Networks identifying pathogens through Raman spectra are vulnerable to unknown inputs, resulting in an uncontrollable false positive rate. To address this issue, first, we developed a novel ensemble of ResNet architectures combined with the attention mechanism which outperforms existing closed-world methods, achieving an accuracy of $87.8 \pm 0.1\%$ compared to the best available model's accuracy of $86.7 \pm 0.4\%$. Second, through the integration of feature regularization by the Objectosphere loss function, our model achieves both high accuracy in identifying known pathogens from the catalog and effectively separates unknown samples drastically reducing the false positive rate. Finally, the proposed feature regularization method during training significantly enhances the performance of out-of-distribution detectors during the inference phase improving the reliability of the detection of unknown classes. Our novel algorithm for Raman spectroscopy enables the detection of unknown, uncatalogued, and emerging pathogens providing the flexibility to adapt to future pathogens that may emerge, and has the potential to improve the reliability of Raman-based solutions in dynamic operating environments where accuracy is critical, such as public safety applications. 

Our model is publicly available: \href{https://github.com/BalytskyiJaroslaw/PathogensRamanOpenSet.git}{https://github.com/BalytskyiJaroslaw/PathogensRamanOpenSet.git}
*Corresponding Author: hr6998@wayne.edu
\end{abstract}

\begin{IEEEkeywords}
Raman spectroscopy; Machine Learning; Pathogen identification in clinical applications; ResNet; Open Set learning; Objectosphere.
\end{IEEEkeywords}

\section{Introduction and Problem Statement}
\label{intro}

Raman spectroscopy involves the scattering of light and its interaction with the chemical bonds present in the material under investigation. This interaction produces a unique spectrum, akin to a fingerprint, that characterizes the material's chemical composition and molecular structure~\cite{pelletier1999analytical}. It was independently discovered in 1928 by Raman~\cite{raman1928new} and Landsberg~\cite{landsberg1928neue} and the appearance of laser spectrometers~\cite{platonenko1964mechanism,hendra1969laser} further expanded its capabilities and applications. Raman spectroscopy is a reliable, sensitive, non-destructive, and versatile analytical technique to determine the chemical composition and molecular structure of complex substances~\cite{lafuente20151}, where it is already used in a number of applications~\cite{orlando2021comprehensive}, while its portability makes it valuable for both laboratory and field applications~\cite{pelletier1999analytical}.  In addition, its unique properties make it a promising tool for biomedical science~\cite{krafft2015many}, including disease diagnosis~\cite{li2014identification,sigurdsson2004detection,ellis2013illuminating,guevara2018use}.

One of the crucial applications of Raman spectroscopy is the identification of bacterial infections, which are responsible for approximately 7 million deaths worldwide each year~\cite{fleischmann2016assessment,deantonio2016epidemiology}. While there are effective methods for detecting pathogenic bacteria, such as enzyme-linked immunosorbent assay (ELISA), polymerase chain reaction (PCR), and sequencing-based approaches, these methods often involve significant time requirements to produce results~\cite{howes2014plasmonic,chung2013magneto,levy2018surviving,chaudhuri2008efns}. Furthermore, clinical diagnostic procedures for identifying specific pathogens often involve time-consuming microbiological culture (up to 48 hours) and antibiotic susceptibility testing (up to 24 hours)~\cite{brusselaers2011rising,sullivan2001effect}. During this waiting period, broad-spectrum antibiotics (BSAbx) are commonly prescribed as a precautionary measure~\cite{brusselaers2011rising,sullivan2001effect}. However, it is important to note that while BSAbx can be life-saving, they should be used judiciously due to their potential side effects and contribution to antibiotic resistance. Excessive use of BSAbx can disrupt the healthy gut microbiome, leading to the overgrowth of pathogens like Candida albicans and Clostridium difficile~\cite{brusselaers2011rising,sullivan2001effect}. Disturbingly, the Centers for Disease Control and Prevention (CDC) has reported that over $30\%$ of patients are treated with antibiotics unnecessarily~\cite{fleming2016prevalence}. The delay in accurately detecting pathogens leads to extended hospital stays, escalated medical expenses, heightened antibiotic resistance, and ultimately, increased mortality rates~\cite{davies2010origins}. To address this issue, Raman spectroscopy offers immense potential as a highly sensitive, culture-free, cost-effective, and rapid identification method. By employing Raman spectroscopy, targeted antibiotics can be administered, thus mitigating the development of antimicrobial resistance~\cite{ho2019rapid}. This approach allows for timely and effective treatment decisions, minimizing the negative impacts associated with delayed pathogen identification.

The application of Raman spectroscopy extends beyond the identification of pathogenic bacteria and encompasses diverse areas such as the diagnosis of COVID-19~\cite{jadhav2021development}, food safety~\cite{neng2020application}, identification of contaminants in pharmaceuticals~\cite{valet2016raman}, and homeland security~\cite{izake2010forensic,choi2019surface,mogilevsky2012raman}. However, in this article, our primary focus is on the identification of pathogenic bacteria using the bacteria-ID dataset~\cite{ho2019rapid}.

To extract meaningful information from Raman spectra, data analysis, and processing are necessary. While manual approaches, such as the ``Ramanome" concept utilizing 31 specific Raman peaks, have been employed~\cite{teng2016label,tao2017metabolic}, they are not sufficiently reliable. This is because spectral information encompasses more complex characteristics beyond these 31 peaks, and inter-class differences pose challenges for manual classification~\cite{klein2019detection,yu2020classification}. Moreover, due to the low probability of Raman scattering, meaningful spectral information can be easily obscured by background noise~\cite{nie1997probing}. Additionally, the large volume of spectral data can be challenging to handle in practical applications, necessitating reliable and efficient quantitative methods facilitated by Machine Learning-driven tools~\cite{rohleder2005comparison}.

We follow the definition of Machine Learning (ML) by Francois Chollet as ``the effort to automate intellectual tasks normally performed by humans"~\cite{chollet2021deep}. The ML framework aims to find a suitable representation of the data, allowing classification rules to be automatically derived rather than hard-coded. ``Deep learning (DL) is a specific subfield of machine learning: a new take on learning representations from data that puts an emphasis on learning successive layers of increasingly meaningful representations"~\cite{chollet2021deep}. In our paper, the above-mentioned layers of data representations are implemented using Deep Neural Networks (DNN). According to~\cite{chollet2021deep}, unlike DL, shallow learning approaches use only one or two consecutive data representation layers.

Shallow learning models, in particular, principal component analysis (PCA) combined with linear discriminant analysis (LDA), are often used to analyze the Raman spectra~\cite{zheng2021serum,liu2016raman,houhou2021comparison,song2020raman,kongklad2022discriminant}. DL models have also been successfully applied to classify molecular spectra~\cite{liu1993chemometric,maruthamuthu2020raman,thrift2020deep,lussier2020deep,peiffer2020machine,lu2020combination} and have shown better performance compared to shallow ones~\cite{gniadecka1997diagnosis}. The vanishing gradient problem~\cite{hochreiter2001gradient} prevents a further boost of the model's performance by a naive approach of 
adding extra layers. ResNet architecture~\cite{he2016deep} fixes this problem by introducing skip connections. ResNet and its modifications have been successfully applied to classify Raman spectra, outperforming shallow models by a large margin, as shown by other authors~\cite{ho2019rapid,balytskyi2022raman,deng2021scale, zhou2022ramannet} and in our previous work~\cite{balytskyi2022raman}.

While deep neural networks (DNNs) excel at identifying classes encountered during the training phase, their behavior becomes unpredictable when confronted with spectra belonging to unknown classes that were not part of the training data, known as out-of-distribution (OOD) samples~\cite{geng2020recent,parmar2023open}. Typically, the SoftMax layer~\cite{bridle1990probabilistic} is used to interpret DNN outputs as probabilities, and the classification result corresponds to the output with the highest SoftMax score. However, as outlined in ~\cite{matan1990handwritten}, even a slight difference in logit values between the winning and runner-up classes can lead to vastly different probability values from the SoftMax layer. Moreover, the SoftMax procedure involves logit value normalization, rendering it inherently closed-world~\cite{geng2020recent} and thus unable to reliably identify OOD samples. Consequently, DNNs often produce incorrect and overly confident predictions when faced with OOD samples. For example, as shown in~\cite{goodfellow2014explaining,nguyen2015deep}, when DNNs encounter ``foolish" and ``rubbish" images visually far from the class from the training catalog, but still produce high confidence scores. Another example is the incorrect and confident classification of a \textit{crab} image as the \textit{clapping} class, despite the fact that no crab-related items were present during training~\cite{drummond2006open}. This necessitates the use of specialized ML techniques capable of identifying OOD samples, as the false positive (FP) rate estimated on large-scale datasets exceeded $70\%$ and, in some cases, was close to $100\%$~\cite{roady2020open,song2020critical}.

The biggest concern in terms of clinical use is that a classifier trained on known classes of bacteria would classify a new type of bacteria as belonging to a known class with high confidence~\cite{ren2019likelihood}. This issue is challenging to mitigate in practice, as it is difficult to anticipate and account for all the potential classes that a classifier might encounter in an unpredictable environment. Some ML systems have been developed to handle this problem, particularly in areas like medical image classification~\cite{guo2022margin}, safety-critical applications~\cite{zhou2023uncertainty}, and environmental monitoring~\cite{morgan2022open}. To tackle this challenge, in our research, we introduce a new ML algorithm designed to accurately identify pathogenic bacteria using Raman spectroscopy in real-world scenarios. This algorithm not only successfully classifies the pathogens already listed in its catalog but also reliably distinguishes and rejects pathogens that are not included in the catalog thereby enhancing patient care and treatment outcomes

Our article is structured as follows. In Section~\ref{split}, we describe the dataset and its division into in-catalog and out-of-catalog samples. In Section~\ref{backbone}, we present our custom ResNet architecture that leverages the strength of the attention mechanism to achieve enhanced performance compared to existing DNN architectures in closed-world scenarios. Furthermore, we highlight the limitations associated with closed-world approaches. In Section~\ref{ENTOBJregularization}, we combine our backbone ResNet architecture with an Entropic Open Set and an Objectosphere loss functions, demonstrating drastic improvement over the naive closed-world approaches in handling the unknowns. To minimize the occurrence of inconclusive outcomes for in-catalog samples, we augment our combined deep neural network with the one-vs-rest classifier in Section~\ref{onevsrest}. In Section~\ref{OODdetectors}, we demonstrate a substantial enhancement in the performance of novelty detectors following the implementation of our proposed feature regularization, as opposed to using naive approaches. Therefore, we demonstrate that our integrated DNN architecture outperforms currently available techniques for both closed- and open-world applications. We present our final conclusions in Section~\ref{Conclusions}. Our previous proof-of-concept work was done in~\cite{balytskyi2022raman}.

\section{Overview of open set learning strategy and splitting the data into $\mathcal{K}/\mathcal{I}/\mathcal{N}$ categories}
\label{split}

In general, current open-set learning approaches belong to two broad classes: generative and discriminative~\cite{geng2020recent}. Generative methods model the joint distribution of input features and labels to estimate the probability that a given sample is OOD, while discriminative methods directly learn the decision boundary between classes to classify the input data based on their features. However, generative methods have been shown to be less efficient than discriminative methods with a well-chosen background class for any but simple datasets~\cite{lee2017training,yu2017open}, so we will focus on discriminative methods in our further considerations.

In the discriminative modeling we consider, we classify the data into three categories: $\mathcal{K}$, $\mathcal{I}$, and $\mathcal{N}$, similar to our previous work~\cite{balytskyi2022raman}. $\mathcal{K}$ corresponds to the known category, the classes of interest that the DNN prioritizes to identify. $\mathcal{I}$ are classes belonging to the background category that the DNN ``ignores" in order to identify $\mathcal{K}$ with greater confidence in a procedure called feature regularization. Finally, $\mathcal{N}$ corresponds to samples corresponding to classes not seen during the training phase of the DNN and which the DNN seeks to distinguish from the $\mathcal{K}$ classes. Only $\mathcal{K}$ and $\mathcal{I}$ are used during the training of the DNN, and the DNN is completely unaware of $\mathcal{N}$ until the testing stage.

For our purposes, we use the bacteria-ID dataset~\cite{ho2019rapid}, which contains 30 pathogen classes shown in Table~\ref{tab:Classes}, with 2000 spectra per class for training, 100 spectra per class for fine-tuning, and 100 spectra per class for testing. To test our ML algorithm in open-world learning settings, we split our dataset into 4 parts, $p_1$, $p_2$, $p_3$, and $p_4$. We assign the pathogen group $p_1$ to $\mathcal{K}$ or ``known classes of interest" since those are extremely common and contagious. Antibiotic-resistant or susceptible pathogens corresponding to the $p_4$ group are particularly harmful to patients and a burden on healthcare systems. Misclassification of these pathogens is extremely problematic, especially if any errors are made in the classification between a susceptible strain of the pathogen and a resistant strain (such as MSSA vs. MRSA). Therefore, we classified the $p_4$ pathogens group as $\mathcal{N}$,  to highlight the ability of our algorithm to identify ``never before seen" samples while keeping high accuracy on the known ones. The $p_2$ and $p_3$ groups are often antibiotics susceptible, but typically found in the body. In our experiments, we tested both options, assigning them as both $\mathcal{K}$ and $\mathcal{I}$ in different runs of the experiments for this proof of concept work.

\begin{table}[]
    \begin{tabular}{|c|l|}
    \hline
      \textbf{    Class    } & \textbf{    Pathogen Name  }\\
      \hline
      $p_1$, used as $\mathcal{K}$ & Group A Strep, Group B Strep, Group C Strep, \\
      
      &
      Group G Strep, E. coli 1, E. coli 2,  E. cloacae, \\
      
      &
      P. mirabilis,  
      S. marcescens,  C. albicans. \\

      \hline
      $p_2$, used as $\mathcal{K/I}$ & E. faecalis 1, E. faecalis 2, E. faecium, \\
      
      &
      
      P. aeruginosa 1,
      P. aeruginosa 2.   \\
 
      \hline
      $p_3$, used as $\mathcal{I}$  & S. Epidermidis, S. lugdunensis, S. sanguinis, \\
      
      &
      K. aerogenes,
       C. glabrata.\\

      \hline
      $p_4$, used as $\mathcal{N}$ & MRSA 1, MRSA 2, MSSA 1, MSSA 2, \\
      &
      MSSA 3,   S. pneumoniae 1, S. pneumoniae 2, \\
      
      &
      K. pneumoniae 1,
      K. pneumoniae 2, S. enterica. \\

      \hline
    \end{tabular}
    \caption{List of pathogen names and their division into categories $\mathcal{K}$, $\mathcal{I}$, and $\mathcal{N}$.}\label{tab:Classes}
\end{table}

 \begin{figure*}[h]{}
  \centering
\includegraphics[width=0.8\textwidth]{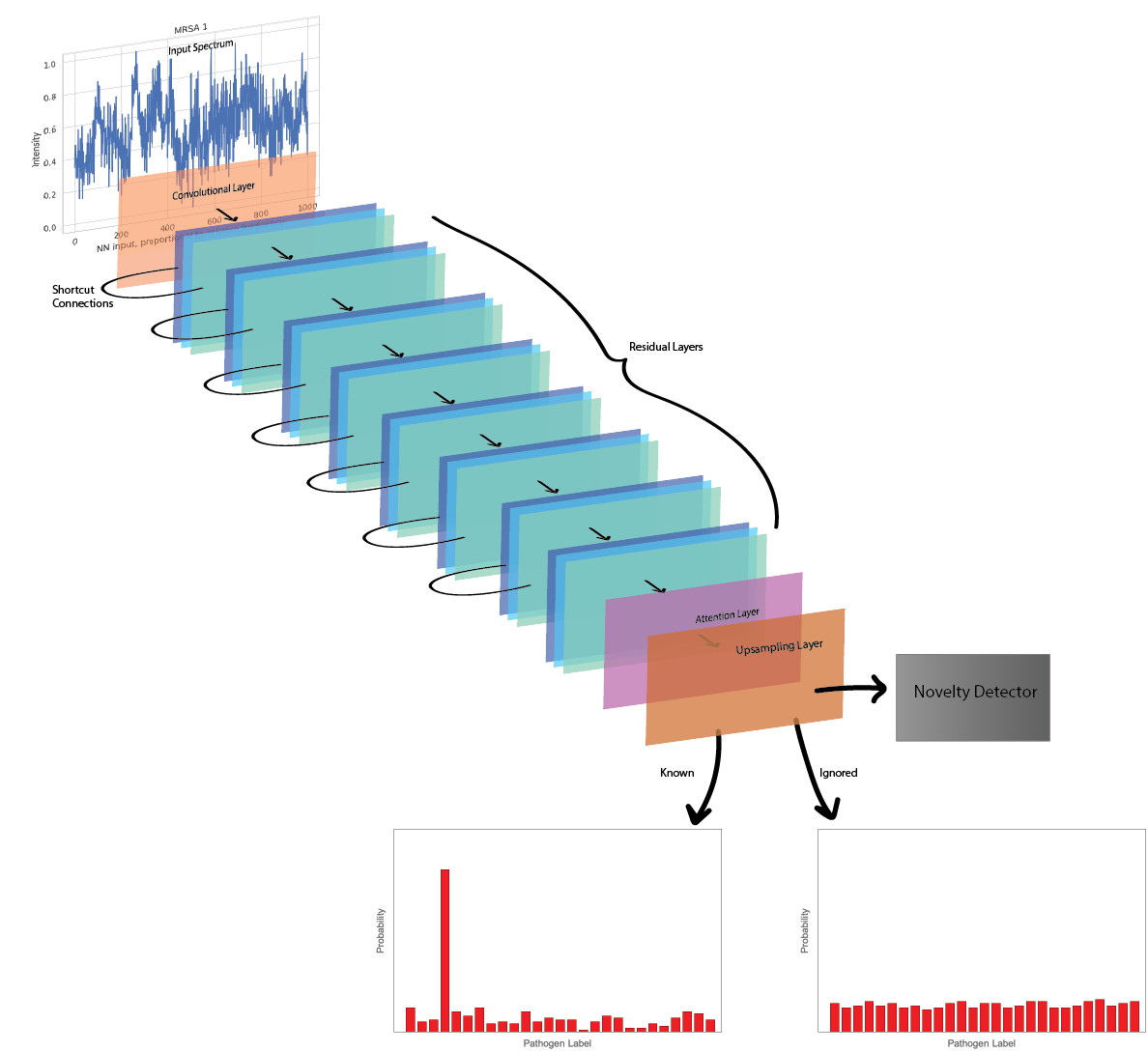}
  \caption{Schematic representation of our custom ResNet architecture. The DNN processes input pathogen Raman spectra into classification probabilities while simultaneously detecting OOD samples. Section~\ref{ENTOBJregularization} describes feature regularization using the $\mathcal{I}$ class, while an additional OOD detector is implemented in Section~\ref{OODdetectors}.  
  }
  \label{CustomResNet}
\end{figure*}

\begin{figure}[]{}
  \centering
 \includegraphics[width=\linewidth]{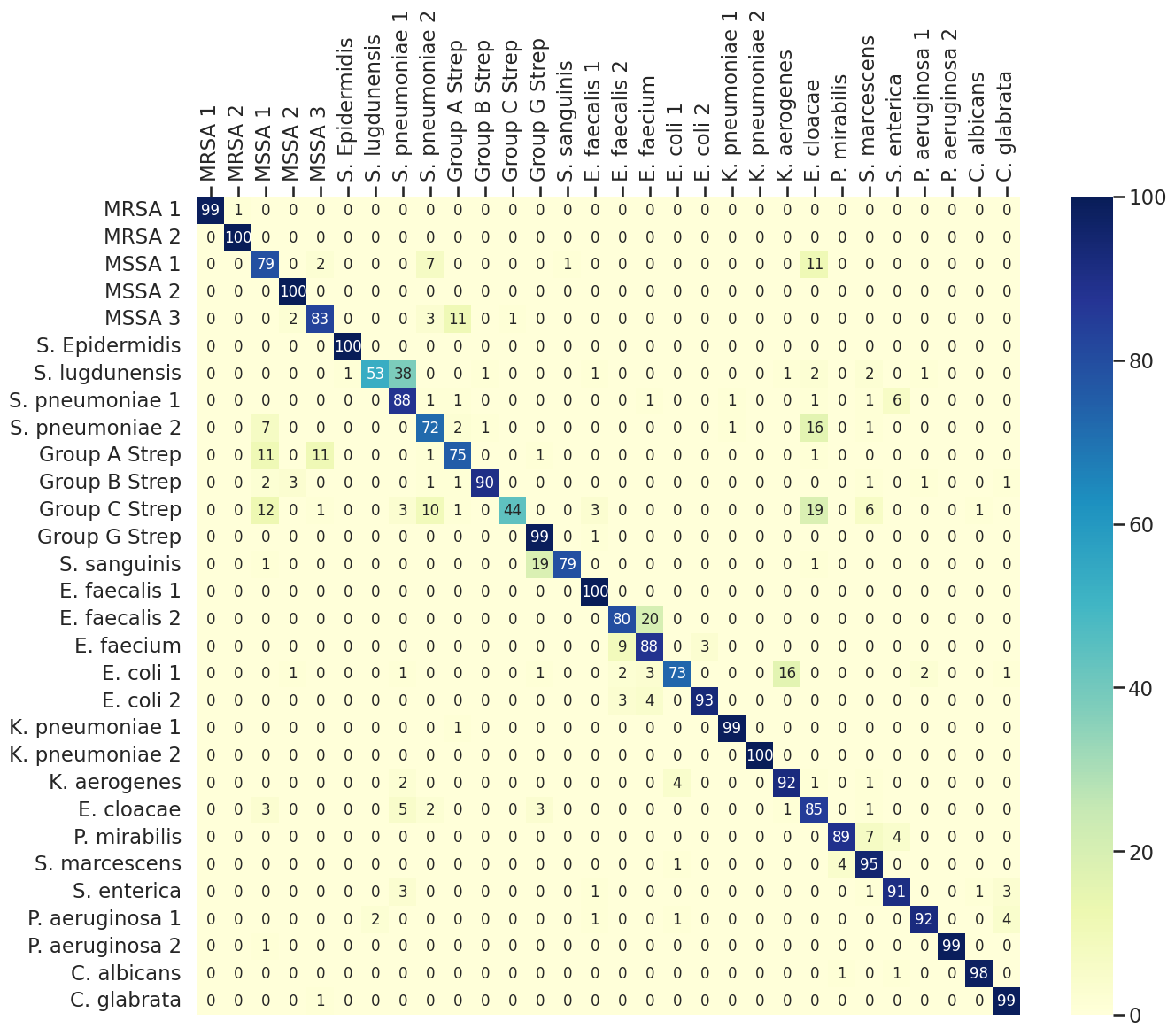}
  \caption{Correlation table for all 30 pathogens in ``closed world" settings when all pathogen classes are known, $\mathcal{K} = p_1 + p_2 + p_3 + p_4$, $\mathcal{I, N} = \emptyset$. Average accuracy = $87.8 \pm 0.1\%$.}
  \label{30Classes}
\end{figure}

However, as mentioned before, it is necessary to carefully assign the pathogen classes to the background category $\mathcal{I}$ in order for the DNN to be efficient. Note that $p_1$ and $p_2$ are closer in their characteristics than $p_1$ and $p_3$. Both groups $p_1$ and $p_2$ consist mainly of streptococcal species and have streptococcal species associated with respiratory and invasive infections. Although groups $p_1$ and $p_3$ also share some common features, such as the presence of Staphylococcus species, in general, $p_1$ and $p_2$ are much closer to each other in their characteristics. Due to the low signal-to-noise ratio of this dataset, it is necessary to keep $\mathcal{K}$ and $\mathcal{I}$ \textit{sufficiently distinct} from each other. 

As we will show in the following Sections, if this condition is not met, the DNN is forced to focus on highly similar samples and “ignore” them simultaneously, which is an inconsistent task and leads to significant performance degradation. The case when $\mathcal{K} = p_1$ and $\mathcal{I} = p_2$ has significantly worse performance compared to all other data partitions, which highlights the importance of choosing $\mathcal{I}$ correctly. It is much more efficient to assign $\mathcal{K} = p_1$ and $\mathcal{I} = p_3$ or $\mathcal{K} = p_1 + p_2$ and $\mathcal{I} = p_3$ as we show further in the text.

\section{Backbone Neural Network Architecture and Limitations of Closed World Approaches} 
\label{backbone}

\begin{figure*}[h]
  \centering
  \includegraphics[width=1.25\textwidth]{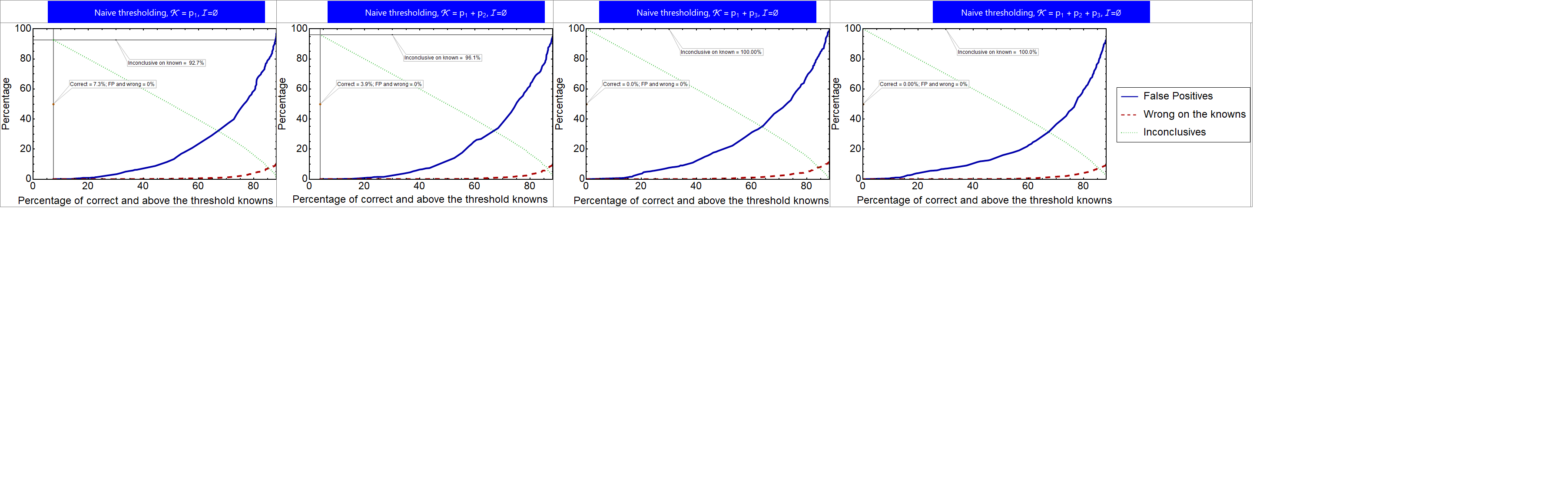}
\vspace{-125pt}

  \caption{False positive rate on $\mathcal{N}$, error and inconclusive rates on $\mathcal{K}$ as a function of correct classification rate for naive approaches.}
  \label{NaiveTHR}
\end{figure*}

We construct our DNN architecture based on custom ResNet architecture, similar to our previous work~\cite{balytskyi2022raman}. However, due to the low signal-to-noise ratio of the current dataset, to boost the accuracy, we supplement our architecture with a squeeze-and-excitation (SE) attention mechanism~\cite{hu2018squeeze} and up-scaling layers. Since this data set is very noisy, we supplement only the last residual block with the SE attention with the goal of enhancing only the high-level features. A schematic representation of our DNN is shown in Fig.~\ref{CustomResNet}. Finally, similar to our previous work, we assembled our ResNets into an ensemble:
\begin{equation}
PredictionEnsemble = \frac{1}{5}\sum_{i = 1}^5 Prediction\left[i\right] 
\end{equation}
To ensure the stability of our model's performance, first, we conducted 20 runs of our model and assessed the accuracy of a single run using all 30 pathogen classes, $\mathcal{K} = p_1 + p_2 + p_3 + p_4$, $\mathcal{I} = \emptyset$. We subsequently grouped these 20 models into 4 ensembles, each consisting of 5 models. As shown in Table~\ref{tab:model-accuracies}, the average accuracy over all 30 classes of an individual model run stands at $87.5\pm 0.4\%$, while the accuracy of the ensemble is $87.8\pm 0.1\%$. Thus, using a model ensemble results in a marked increase in accuracy and reduction in variance. The corresponding correlation table is shown in Fig.~\ref{30Classes}. The ensemble architecture we propose surpasses the current state-of-the-art closed-world DNNs, which achieve accuracies of $82.2 \pm 0.3\%$\cite{ho2019rapid}, $84.7 \pm 0.3\%$\cite{zhou2022ramannet}, and $86.7\pm 0.4\%$~\cite{deng2021scale}, respectively. This remarkable performance motivates us to utilize an ensemble of ResNet models augmented with an SE attention mechanism not only in the closed-world setting but also in subsequent open-world settings considered in the next Sections.

\begin{table}[h]
    \centering
    \caption{Single runs and ensemble accuracies on all 30 classes, $\mathcal{K} = p_1 + p_2 + p_3 + p_4$, $\mathcal{I} = \emptyset$.}
    \begin{tabular}{|p{0.9cm}|c|c|}
        \toprule
        Runs \# & Accuracy of a run & Ensemble accuracy  \\
        \midrule
        1 - 5 & 87.6\%, 87.2\%, 88.2\%, 87.7\%, 87.4\%   &  87.8\% \\
        \hline
        6 - 10 & 88.3\%, 87.9\%, 87.4\%, 87.8\%, 87.7\% &  87.9\%  \\
        \hline
        11 - 15 & 87.7\%, 87.5\%, 86.9\%, 87.3\%, 87.1\% &  87.9\%  \\
        \hline
        16 - 20 & 86.7\%, 87.5\%, 87.2\%, 87.4\%, 87.1\% &  87.6\%  \\
        \midrule
        Accuracy & $87.5 \pm 0.4\%$ & $87.8 \pm 0.1\%$ \ \\
        \bottomrule
    \end{tabular}
    \label{tab:model-accuracies}
\end{table}

\begin{figure}[h]{}
  \includegraphics[width=\linewidth]{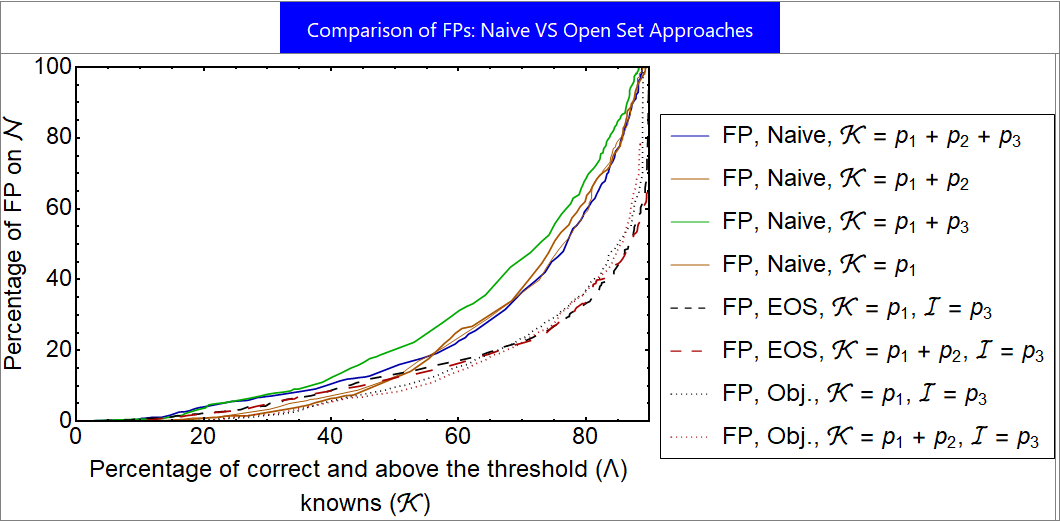}
  \caption{Comparison of false positive rates by naive vs Open Set methods.}
  \label{NaiveVSOpenSet}
\end{figure}

Before we get into the open-world settings, we need to establish the required definitions. The input of the DNN $x$ corresponds to the intensity values of the input Raman scattering spectra, and the corresponding output represents the probability of the spectrum belonging to a specific class of pathogens $p$, given by the logit values $l_p\left(x\right)$. The logit values are obtained by multiplying the output  from the second to last layer of the DNN, called the deep features $F\left(x\right)$, by the weights $W$, and the probability of a spectrum belonging to a particular class of pathogens $p$ is obtained from ``softmaxing" procedure, defined as:
\begin{equation}\label{Softmaxing}
    l_p(x) = W \cdot F\left(x\right), \ S_p\left(x\right) = \frac{e^{l_p\left(x\right)}}{\sum_{p} e^{l_p\left(x\right)}}
\end{equation}
The resulting value is in the interval $S_p\left(x\right) \in \left[0, 1\right]$ with $\sum_p S_p\left(x\right) = 1$, and thus $S_p\left(x\right)$ can be interpreted as a probability measure.

In the case where the input belongs to the category $\mathcal{K}$, it is classified as the pathogen that has the highest softmax score in Eqn~\ref{Softmaxing}. In the perfect case scenario, the DNN's output from the sample belonging to i-th class in the pathogen catalog should return: 

\begin{equation}
\left(Pathogen \ class \in \mathcal{K}\right)\left[i\right] \rightarrow \underbrace{\left[0, \cdots,\underbrace{1}_{i-th \ position}, \cdots, 0 \right]}_{Length = \lvert\mathcal{K}\lvert},
\end{equation}
where $\lvert\mathcal{K}\lvert$ represents the number of pathogen classes in the $\mathcal{K}$ catalog.

The simplest way to handle category $\mathcal{N}$ is to threshold the softmax score~\cite{matan1990handwritten,de2000reject,fumera2002support}. The main assumption of this approach is that the categories $\mathcal{K}$ and $\mathcal{N}$ are sufficiently separated in the feature space and the DNN's output corresponding to $\mathcal{N}$ approaches:

\begin{equation}
\mathcal{N} \rightarrow \underbrace{\left[\frac{1}{\lvert\mathcal{K}\lvert}, \cdots, \frac{1}{\lvert\mathcal{K}\lvert} \right]}_{Length = \lvert\mathcal{K}\lvert},
\end{equation}
and Shannon entropy~\cite{shannon1948mathematical} 
reaching its maximum value $\log_2\left(\lvert\mathcal{K}\lvert\right)$.

In case this condition is fulfilled, it is possible to introduce the threshold $\Lambda$ such that $\mathcal{K}$ and $\mathcal{N}$ are separated as $max\left(\mathcal{N}\right) < \Lambda$ while $max\left(\mathcal{K}\right) > \Lambda$. In practice, if the maximum value of the softmax score is less than $\Lambda$, this is classified as an inconclusive ``I don't know what it is" result, which may mean that the sample belongs to category $\mathcal{N}$ outside the $\mathcal{K}$ catalog. Another possibility is that the sample belongs to $\mathcal{K}$, but with low confidence. Thus, our goal is to separate samples outside the catalog $\mathcal{N}$ while minimizing the number of inconclusive results for samples in the catalog.

Since the correct classification, error, and inconclusive rates on $\mathcal{K}$ as well as the FP rate on $\mathcal{N}$ are all functions of the global threshold $\Lambda$, it is convenient to represent the FP, error, and inconclusive rates as a function of the correct classification rate, and we plot the results corresponding the naive threshold approaches in Fig.~\ref{NaiveTHR}. A striking feature can be observed: the FP rate for unknowns is much higher than the error probability for knowns and can be close to $100\%$, so special methods are needed to solve this problem, which will be implemented in the following Sections. In practice, as Fig.~\ref{NaiveTHR} demonstrates, $\mathcal{K}$ and $\mathcal{N}$ are not sufficiently separated, resulting in a high FP rate, and therefore the assumption discussed above is false. In the next Section, we show that by introducing an additional ``ignored" category $\mathcal{I}$ and Open Set methods, this problem can be mitigated, as illustrated in Fig.~\ref{NaiveVSOpenSet}. The solid lines corresponding to the FP rates of the naive approaches are much higher than the dotted and dashed lines corresponding to the Open Set methods labeled ``EOS" and ``Obj." and discussed further.

\begin{figure}[]{}
  \includegraphics[width=\linewidth]{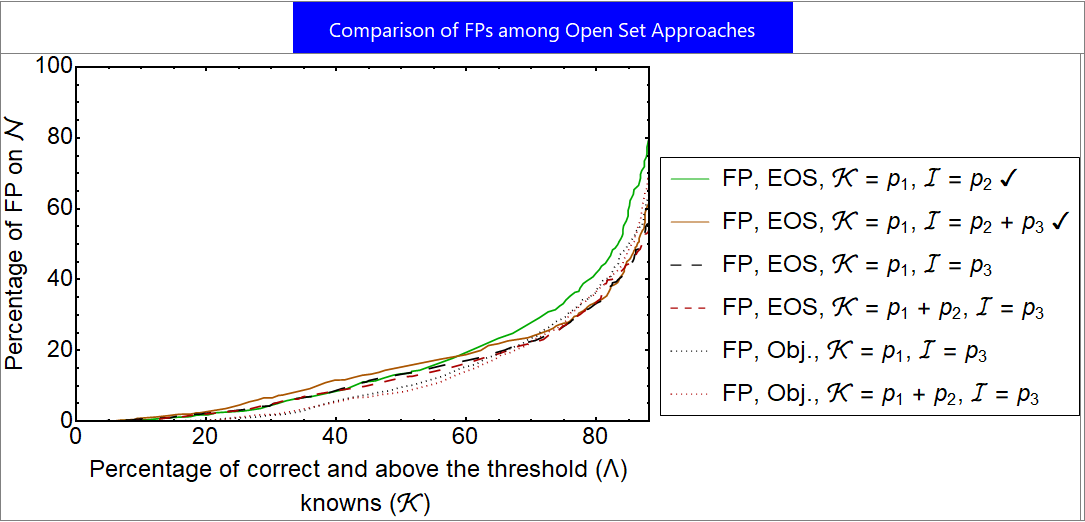}
  \caption{Comparison of FP rates of all  Entropic Open Set (EOS) and Objectosphere (Obj.) experiments. A noticeably higher FP rate when $\mathcal{I}$ dataset includes $p_2$ can be observed, shown by the $\checkmark$ mark.}
  \label{CompareFPOpen}
\end{figure}

\section{Feature regularization by Entropic Open Set and Objectosphere methods}
\label{ENTOBJregularization}

In order to improve separation between $\mathcal{K}$ and $\mathcal{N}$, we introduce the ``ignored" category, $\mathcal{I}$. This approach was originally proposed in~\cite{dhamija2018reducing} and has been shown to be highly efficient for open-world Raman spectroscopy purposes~\cite{balytskyi2022raman}. 

\begin{figure*}[]{}
  \centering
  \includegraphics[width=\linewidth]{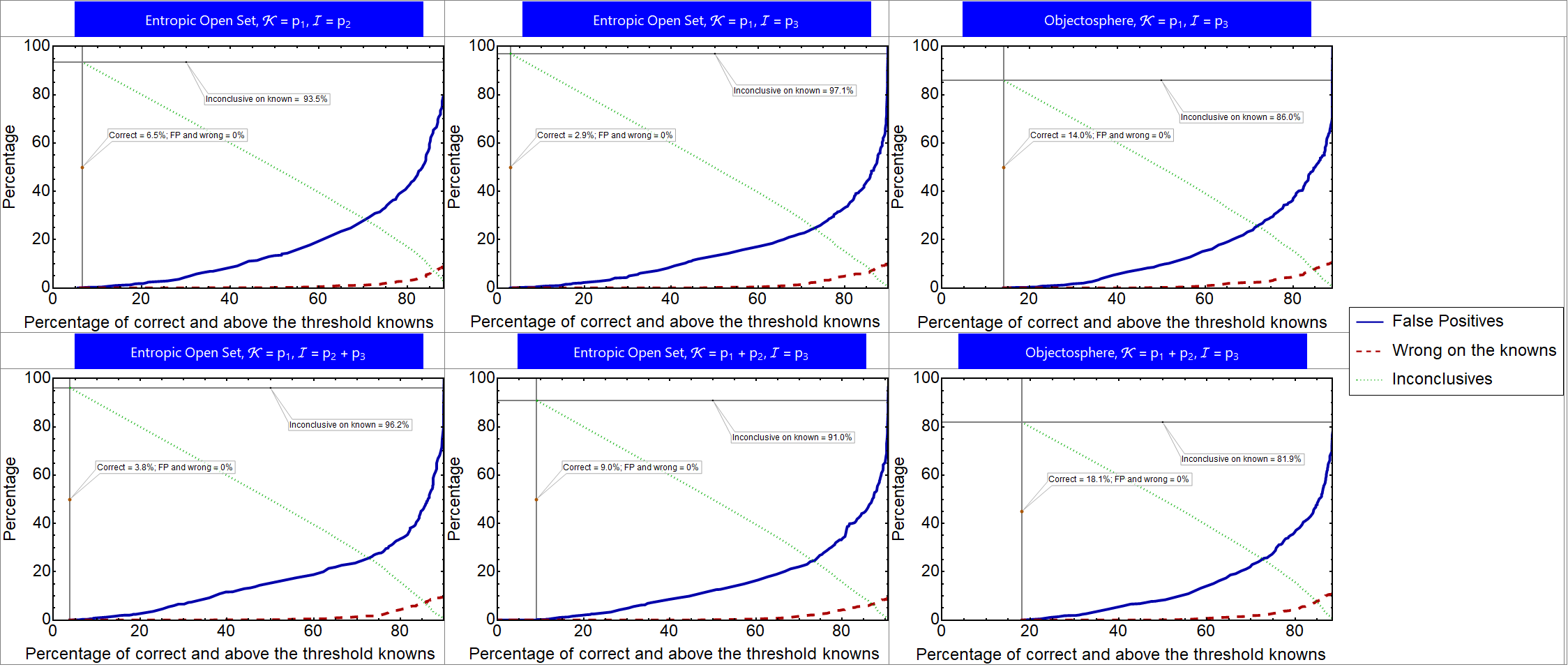}
  \caption{FP rate on $\mathcal{N}$, error and inconclusive rates on $\mathcal{K}$ as a function of correct classification rate for Entropic Open Set (EOS) and Objectosphere (Obj.) approaches.}
  \label{Combined_ENT_OBJ}
\end{figure*}

This method consists modifying the loss function during the training, and consists of two parts. First, the Entropic Open Set (EOS) loss function~\cite{dhamija2018reducing} is defined as: 
\begin{equation}
V_E\left(x\right) = \begin{cases}
- \log\left(S_p\left(x\right)\right), if \ x \in\mathcal{K} \\
-\frac{1}{\lvert\mathcal{K}\lvert}\sum_{p = 1}^{\lvert\mathcal{K}\lvert} \log\left(S_p\left(x\right)\right), if \ x \in\mathcal{I}
\end{cases},
\end{equation}
thus, for category $\mathcal{K}$, it reduces to the usual categorical cross-entropy loss function, while for the case where $x\in \mathcal{I}$, $V_E\left(x\right)$ aims to maximize the Shannon entropy and uniformly distribute the output of the DNN over the knowns.

Second, in general, classes belonging to $\mathcal{K}$ tend to have higher absolute values of deep features $\lvert\lvert F\left(x\right) \rvert\rvert$ than classes belonging to $\mathcal{N}$. Thus the Objectosphere loss function aims to increase this separation by using the deep feature $F\left(x\right)$ parameter as:  

\begin{equation}
V_O\left(x\right) = V_E\left(x\right) + \alpha
\begin{cases}
max\left(\beta - \lvert\lvert F\left(x\right) \rvert\rvert^2, 0 \right), if \ x \in\mathcal{K} \\
\lvert\lvert F\left(x\right) \rvert\rvert^2, if \ x \in\mathcal{I}
\end{cases},
\label{OBJ_loss_function}
\end{equation}
where the values of $\alpha$ and $\beta$ are adjusted to minimize the inconclusive rate on the $\mathcal{K}$ category  by the model cross-validation, and $\lvert\lvert\cdot\lvert\lvert$ is a regular Euclidean norm. This leads to a minimization of the FP rate on $\mathcal{I}$, and this property generalizes to the $\mathcal{N}$ category, even though DNN is unaware of $\mathcal{N}$ until the testing phase.

The corresponding FP rates on $\mathcal{N}$, error and inconclusive rates on $\mathcal{K}$ as a function of correct classification rate for the Entropic Open Set (EOS) and Objectosphere (Obj.) approaches with different choices of $\mathcal{K}$ and $\mathcal{I}$ are shown in Fig.~\ref{Combined_ENT_OBJ}. As mentioned earlier in Section~\ref{split}, it is crucial to choose the right dataset for the $\mathcal{I}$ category. One can increase the global threshold $\Lambda$ and reduce the FP frequency as well as the error rate to zero at the cost of increasing the frequency of inconclusive results on $\mathcal{K}$. The FP rates of all Open Set learning experiments are plotted in Fig.~\ref{CompareFPOpen}, and one can observe a noticeably higher FP rate when $\mathcal{I}$ includes $p_2$  showed by solid lines and marked by the $\checkmark$ sign.

Although the Open Set Learning methods described above significantly improve the DNN's performance in the open world as shown in Fig.~\ref{NaiveVSOpenSet}, there are a significant number of inconclusive results on $\mathcal{K}$. For example, as shown in Fig.~\ref{Combined_ENT_OBJ}, if the threshold is increased so that both the FP and the error rates are zero, the highest rate of conclusive result achieved is around $18\%$. Therefore, in the next Section, we combine the Open Set approaches implemented here with a one-versus-the-rest classifier to increase the number of conclusive results on $\mathcal{K}$ while keeping the FP and error rate zero.

\section{Combination with the one-vs-rest classifier}
\label{onevsrest}

To reduce the number of inconclusive results, instead of the global  $\left(\Lambda\right)$, we introduce the per-class threshold $\Lambda^\prime$ to classify the DNN's output $S\left( x \right)$:
$$
S\left( x \right) =  \underbrace{\left[s_1, \cdots, s_{\lvert\mathcal{K}\lvert} \right]}_{Length = \lvert\mathcal{K}\lvert}, \
\Lambda^\prime = \underbrace{\left[\lambda_1, \cdots,\underbrace{\lambda_i}_{i-th \ position}, \cdots, \lambda_{\lvert\mathcal{K}\lvert} \right]}_{Length = \lvert\mathcal{K}\lvert}.
$$
If the maximum value of the output exceeds the threshold value for the class, $\underset{k \in \mathcal{K}}{max}\left(s_k\right) = s_i > \lambda_i$, the spectrum is classified as belonging to $\left(Pathogen \ class \in \mathcal{K}\right)\left[i\right]$. Otherwise, in the case $s_i < \lambda_i, i \in \mathcal{K}$, the result is inconclusive.

Since different classes have different rates of confidence represented by the softmax score, the class-adaptive threshold leads to a higher \textit{average} rate of conclusive outcomes. Increasing the per-class threshold reduces both the FP and error rates while increasing the number of inconclusive outcomes, and we compute  $\langle \lambda  \rangle^{FP/ Err. = 0\%}_{\mathcal{K}}$ providing both \textit{FP = 0\%} and \textit{error rate = 0\%}. A similar approach was successfully used for open-world text classification as a part of DOC model~\cite{shu2017doc}.

\begin{figure}[]{}
  \includegraphics[width=\linewidth]{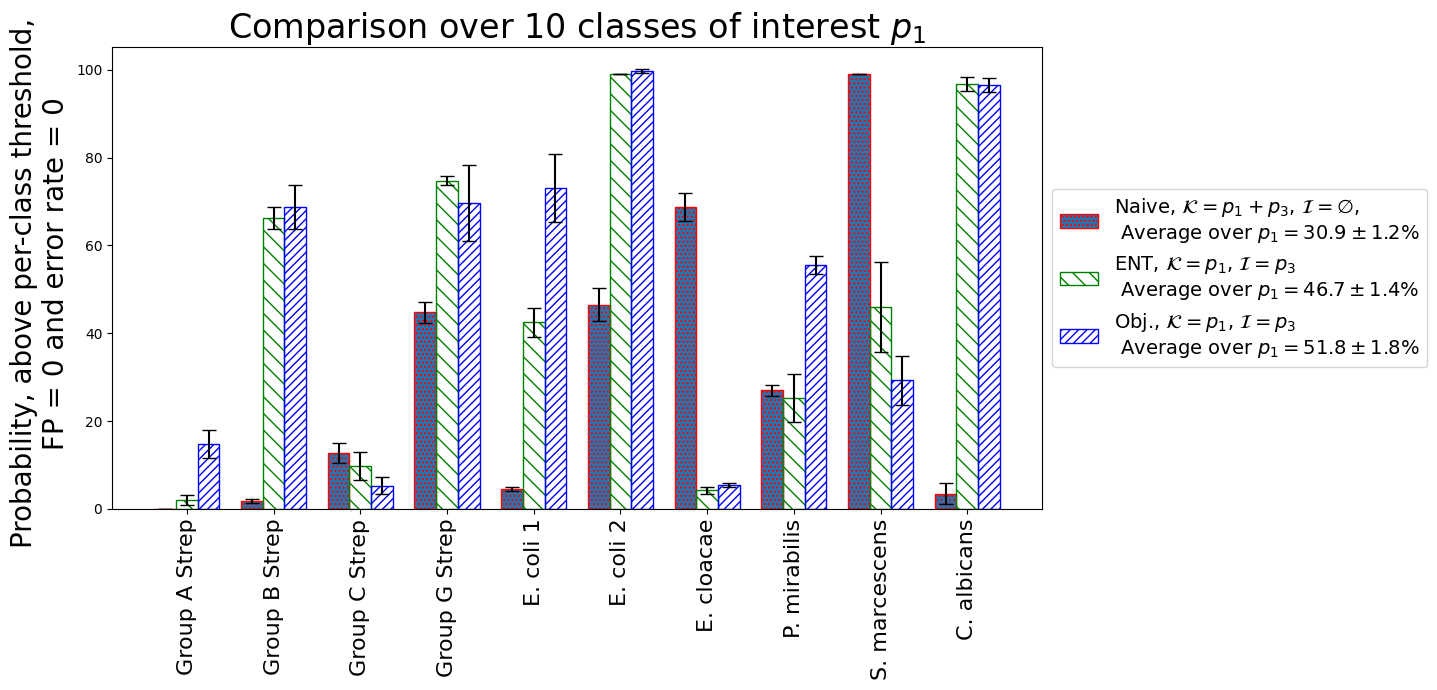}
  \caption{Comparison of conclusive outcomes over classes $\mathcal{K} = p_1$ for naive and Open Set methods. Error bars represent one standard deviation over 4 ensembles.}
  \label{BarChart1}
\end{figure}

\begin{figure}[]{}
  \includegraphics[width=\linewidth]{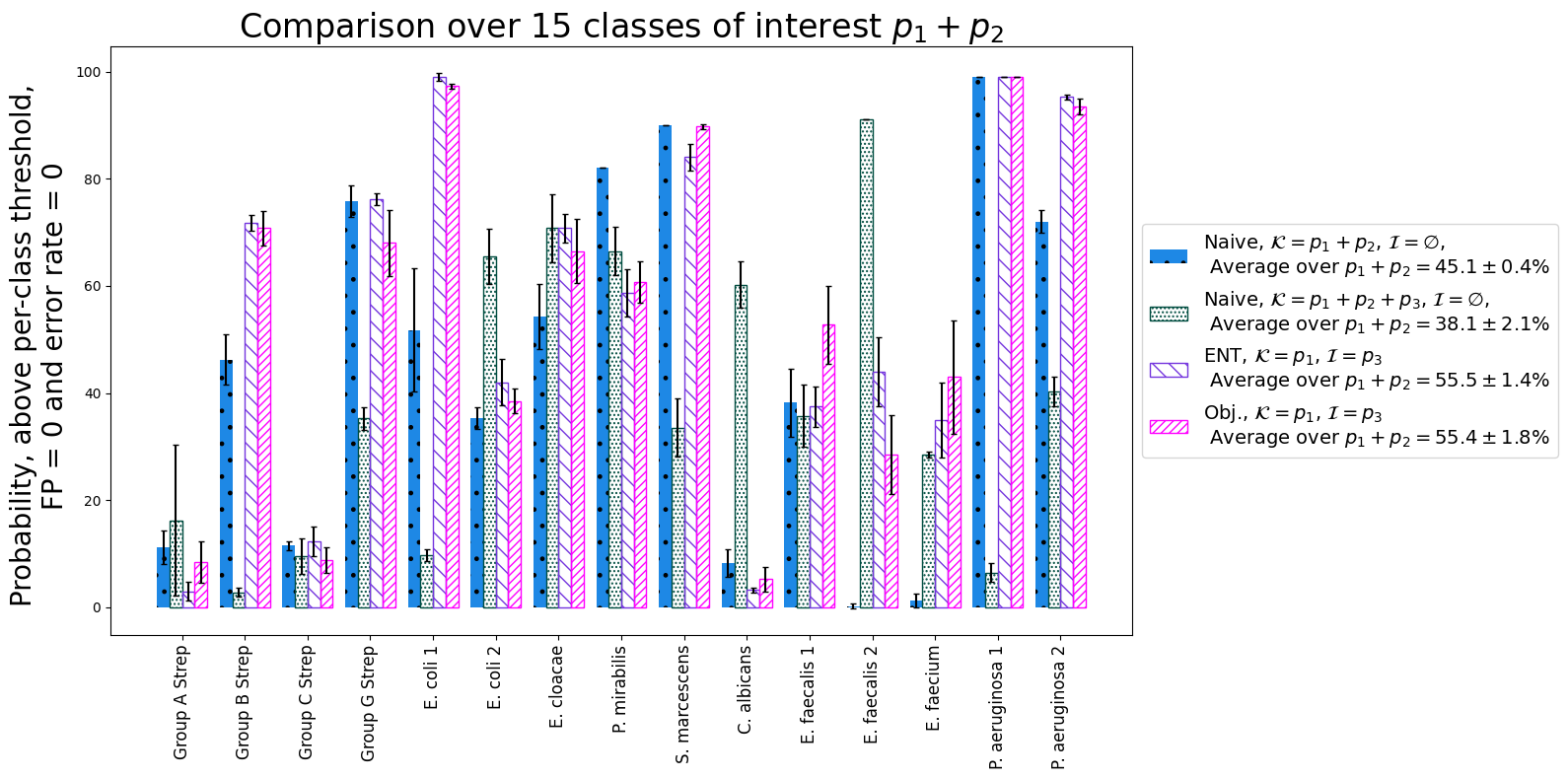}
  \caption{Comparison of conclusive outcomes over classes $\mathcal{K} = p_1 + p_2$ for naive and Open Set methods. Error bars represent one standard deviation over 4 ensembles.}
  \label{BarChart2}
\end{figure}

 \begin{table}[h!]
  \centering
  \begin{tabular}{ |p{2.5cm}||p{1.38cm}|p{1.38cm}|p{1.38cm}| }
    \hline
    & $\langle \lambda  \rangle^{FP/ Err. = 0\%}_{p_1}$ & $\langle \lambda  \rangle^{FP/ Err. = 0\%}_{p_1 + p_2}$ & $\langle \lambda  \rangle^{FP/ Err. = 0\%}_{p_1 + p_2 + p_3}$ \\
    \hline
    Naive: $\mathcal{K} = p_1$, \ \  $\mathcal{I} = \emptyset$ & $40.3 \pm 1.0\%$  & N/A  & N/A \\
    Naive: $\mathcal{K} = p_1 + p_2$, $\mathcal{I} = \emptyset$ & $46.7 \pm 0.8\%$  & $45.1 \pm 0.4\%$  & N/A \\
    EOS: $\mathcal{K} = p_1$, \ \ \ \ $\mathcal{I} = p_2$ $\checkmark$
 & $42.3 \pm  1.0 \%$  & N/A  & N/A \\
    \hline
    Naive: $\mathcal{K} = p_1 + p_3$, $\mathcal{I} = \emptyset$ & $30.9 \pm 1.2 \%$ &   N/A   & N/A \\
    EOS: $\mathcal{K} = p_1$, \ \ \ \ $\mathcal{I} = p_3$& $46.7 \pm 1.4\%$ & N/A  & N/A \\
    Obj.: $\mathcal{K} = p_1$, \ \ \ \ $\mathcal{I} = p_3$ & $51.8 \pm 1.8 \%$   & N/A & N/A\\
    \hline
    Naive: $\mathcal{K} = p_1 + p_2 + p_3$, $\mathcal{I} = \emptyset$ & $37.0 \pm 2.6 \%$ & $38.1 \pm 2.1 \%$  & $44.6 \pm 1.4 \%$ \\
    EOS: $\mathcal{K} = p_1$, \ \ \ \  $\mathcal{I} = p_2+p_3$$ \checkmark$   & $48.1 \pm 1.2 \%$  & N/A & N/A \\
    EOS: $\mathcal{K} = p_1 + p_2$, $\mathcal{I} = p_3$  & $52.1 \pm 1.0\%$ & $55.5 \pm 1.4\%$  & N/A \\
    Obj.: $\mathcal{K} = p_1 + p_2$, $\mathcal{I} = p_3$  & $51.4 \pm 1.5 \%$ & $55.4 \pm 1.8 \%$ & N/A\\
    \hline
  \end{tabular}
  \caption{Comparison of the average rate of conclusive results over the $\mathcal{K}$ category by naive and open set methods.}
  \label{tab:lambdas}
\end{table}
The corresponding results are provided in Table~\ref{tab:lambdas}, and one can observe that when the category $\mathcal{I}$ is chosen appropriately, the Entropic Open Set and Objectosphere approaches consistently outperform the naive thresholding. For example, naive threshoding with $\mathcal{K} = p_1 + p_2 + p_3$  and $\mathcal{I} = \emptyset$ have $\langle \lambda  \rangle^{FP/ Err. = 0\%}_{p_1 + p_2} = 38.1 \pm 2.1\%$, naive threshoding with $\mathcal{K} = p_1 + p_2$  and $\mathcal{I} = \emptyset$ have $\langle \lambda  \rangle^{FP/ Err. = 0\%}_{p_1 + p_2} = 45.1 \pm 0.4\%$, while EOS and Obj. with $\mathcal{K} = p_1 + p_2$ and $\mathcal{I} = p_3$ have $55.5\pm 1.4\%$ and $55.4\pm 1.8\%$ respectively. However, as mentioned before, when $p_2$ is included in the $\mathcal{I}$ dataset, this leads to degradation of performance, as marked by the $\checkmark$ sign in same Table~\ref{tab:lambdas}.

A comparison of the conclusive results for different classes of pathogens is presented in Figs.~\ref{BarChart1} and~\ref{BarChart2}. One can observe that while Entropic Open Set and Objectosphere approaches have a higher rate of conclusive results \textit{on average} over the classes of interest, there are pathogen classes on which naive thresholding has a higher rate of conclusive results.

 \begin{figure*}[]{}
  \centering
 \includegraphics[width=\linewidth]{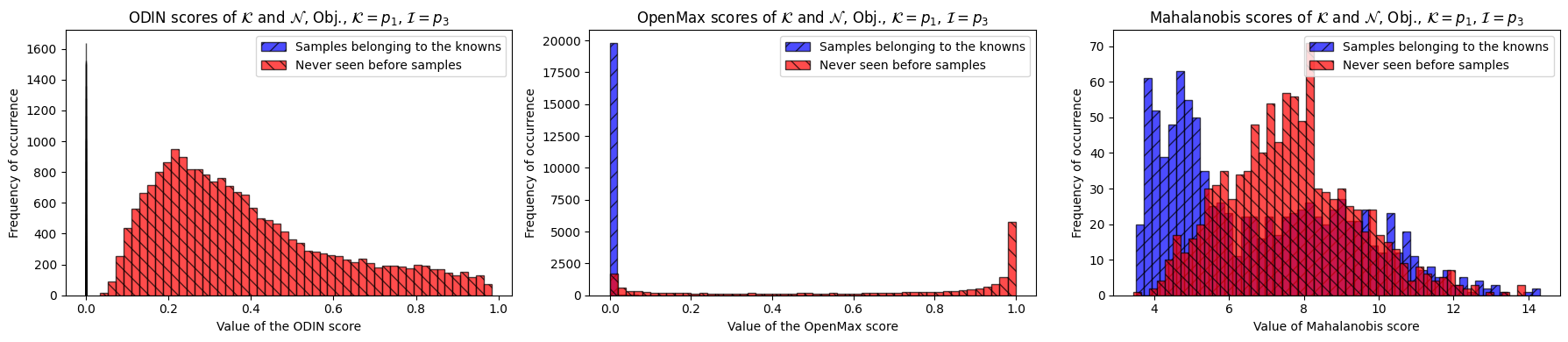}
  \caption{Comparison of OOD detectors after the model training with Objectosphere loss function with $\mathcal{K} = p_1$ and $\mathcal{I} = p_3$. A significantly better $\mathcal{K}/\mathcal{N}$ separation can be observed for the ODIN compared to the OpenMax and Mahalanobis.}
  \label{FirstAllOOD}
\end{figure*}

 \begin{figure*}[]{}
  \centering
 \includegraphics[width=\linewidth]{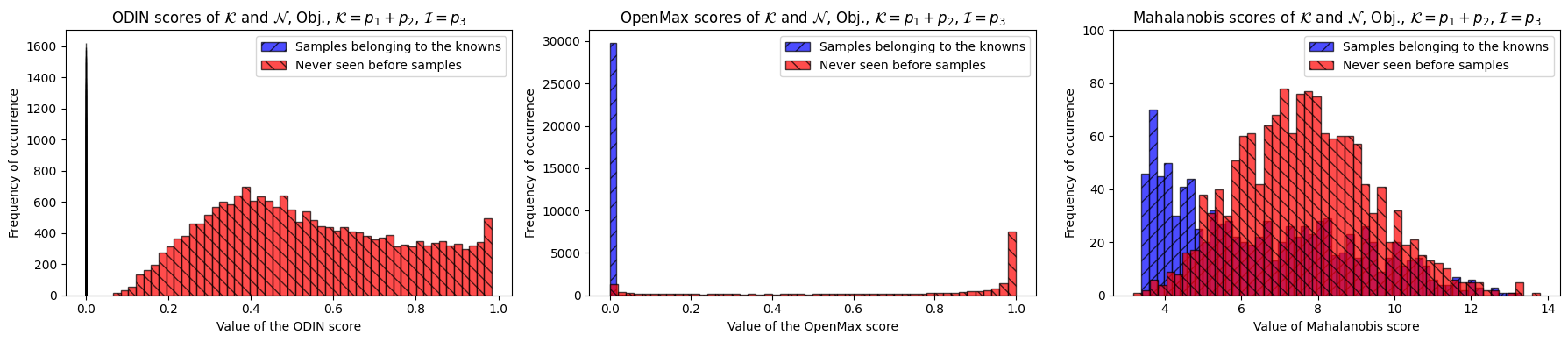}
  \caption{Comparison of OOD detectors after the model training with Objectosphere loss function with $\mathcal{K} = p_1 + p_2$ and $\mathcal{I} = p_3$. Again,  significantly better $\mathcal{K}/\mathcal{N}$ separation can be observed for the ODIN detector.}
  \label{SecondAllOOD}
\end{figure*}

 \begin{figure*}[]{}
  \centering
 \includegraphics[width=\linewidth]{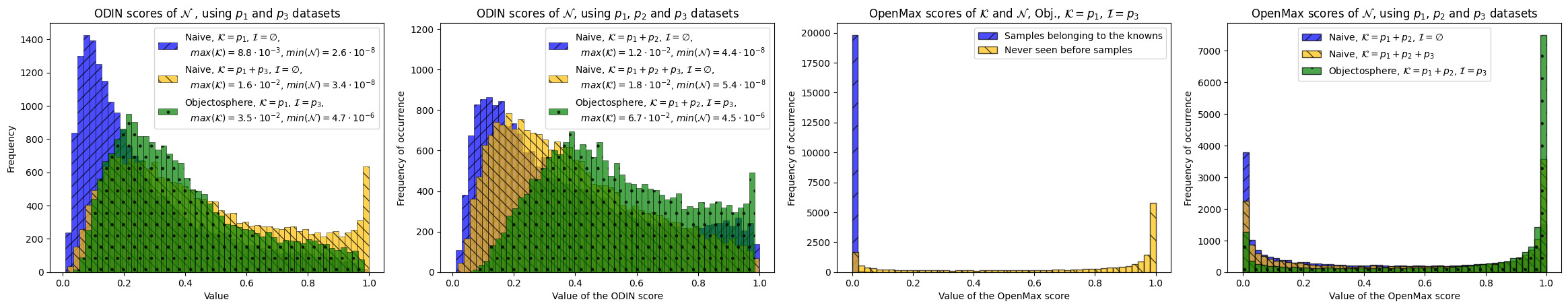}
  \caption{Comparison of ODIN and OpenMax detectors in combination with naive approaches and Objectosphere. One can observe a significantly better $\mathcal{K}/\mathcal{N}$ for the case of Objectosphere. }
  \label{ImprovedOOD}
\end{figure*}

\section{Supplement of OOD detectors and their evaluation}
\label{OODdetectors}

Finally, we implement and test OOD detectors that aim to separate $\mathcal{N}$ samples from $\mathcal{K}$ \textit{after} training, namely Mahalanobis~\cite{lee2018simple}, OpenMax~\cite{bendale2016towards}, and ODIN~\cite{liang2017enhancing} detectors.

As shown in Fig.~\ref{FirstAllOOD} and Fig.~\ref{SecondAllOOD}, the ODIN detector based on input perturbations and temperature scaling performs much better than the other two OOD detectors and separates $\mathcal{K}$ and $\mathcal{I}$ with a significant margin while the other two detectors have a significant $\mathcal{K}/\mathcal{I}$ overlap, similarly to the findings of~\cite{roady2020open}.

At the same time, as shown in Fig.~\ref{ImprovedOOD}, the feature regularization by the Objectosphere loss function during the training boost the ODIN performance even more and leads to a significantly larger margin separating   $\mathcal{K}$ and $\mathcal{I}$ scores. For the Objectosphere with $\mathcal{K} = p_1$ and $\mathcal{I} = p_3$, the maximum value of the ODIN score for known classes is $max\left(ODIN\left(\mathcal{K}\right)\right) = 4.7\cdot 10^{-6}$ and the minimum value of the ODIN score for never seen before classes is $max\left(ODIN\left(\mathcal{N}\right)\right) = 3.5\cdot 10^{-2}$ while for naive thresholding with $\mathcal{K} = p_1 + p_3$ the corresponding values are $max\left(ODIN\left(\mathcal{K}\right)\right) = 3.4\cdot 10^{-8}$ and $max\left(ODIN\left(\mathcal{N}\right)\right) = 1.6\cdot 10^{-2}$. For the case of Objectosphere with $\mathcal{K} = p_1  + p_2$ and $\mathcal{I} = p_3$, the $\mathcal{K}/\mathcal{I}$ margin is even larger, the maximum value of the ODIN score for known classes is $max\left(ODIN\left(\mathcal{K}\right)\right) = 4.5\cdot 10^{-6}$ and the minimum value of the ODIN score for never seen before classes is $max\left(ODIN\left(\mathcal{N}\right)\right) = 6.7\cdot 10^{-2}$, in comparison with naive thresholding for $\mathcal{K} = p_1 + p_2 + p_3$ the corresponding values being $max\left(ODIN\left(\mathcal{K}\right)\right) = 5.4\cdot 10^{-8}$ and $max\left(ODIN\left(\mathcal{N}\right)\right) = 1.8\cdot 10^{-2}$. Additionally, as one can observe from Fig.~\ref{ImprovedOOD}, the histogram corresponding to Objectosphere is noticeably shifted towards larger values of the ODIN scores. Similarly, even though the OpenMax detector performs worse than ODIN, the histograms corresponding to Objectosphere are shifted towards larger values as well. 

If during the training stage, in addition to focusing on the knowns, the DNN has its features regularized by means of the Objectosphere loss function in Eqn.~\ref{OBJ_loss_function}, it leads to a significantly improved separation between knowns and unknowns in comparison with the application of the OOD detector with the naive approaches alone leading to improved reliability of inference.

\section{Conclusions and future work}
\label{Conclusions}

Machine Learning-enabled Raman spectroscopy holds significant promise as a label-free, accurate, and rapid method for identifying pathogens and hazardous contaminants, contributing to the preservation of human lives. However, the reliability and robustness of ML models used in such applications pose limitations, particularly in critical scenarios where complete knowledge of all possible classes cannot be assumed and when there are substantial disparities between test and training data. To address this gap, we developed a unified approach that addresses the problem of reliable and robust classification of open-world Raman spectra by leveraging the capabilities of ResNet combined with the SE attention mechanism and Objectosphere loss function. We evaluated the proposed method on the bacteria-ID database and demonstrated its superiority over existing state-of-the-art methods in both closed and open-world settings. Combination with the one-vs-rest classifier significantly improves the number of inconclusive outcomes while keeping the FP and error rate zero. Additionally, we showed that the conjunction of OOD detectors with our architecture boosts their performance and found that the ODIN detector performs significantly better than the Mahalanobis and OpenMax detectors, making it a valuable supplement for OOD detection. In the future, we aim to adapt our ML algorithm to cater to other critical applications such as public safety and environmental monitoring, benefiting from the adaptability of our proposed model to analyze Raman spectra in diverse contexts.

\section*{Acknowledgments}

This work was supported by funding from the UCCS Cyber Seed Grant as well as the UCCS BioFrontiers Center. Y.B. appreciates fruitful discussions with Drs. Tristan Paul, Guy Hagen, and Sang-Yoon Chang.

\bibliographystyle{IEEEtran}
\bibliography{apssamp.bib}

\end{document}